\documentclass[aps,superscriptaddress,twocolumn,showpacs]{revtex4}
\usepackage{graphicx}

\begin{document}

\title{Surface location of sodium atoms attached to $^3$He nanodroplets}

\author{F. Stienkemeier}
\affiliation{Fakult\"at f\"ur Physik, Universit\"at Bielefeld, D-33615
Bielefeld, Germany}

\author{O. B\"unermann}
\affiliation{Fakult\"at f\"ur Physik, Universit\"at Bielefeld, D-33615
Bielefeld, Germany}

\author{R. Mayol}
\affiliation{Departament E.C.M., Facultat de F\'{\i}sica, Universitat
de Barcelona, E-08028, Spain}

\author{F. Ancilotto}
\affiliation{INFM (Udr Padova and DEMOCRITOS National Simulation
Center, Trieste, Italy) and Dipartimento di Fisica ``G. Galilei'',
Universit\`a di Padova, via Marzolo 8, I-35131 Padova, Italy}

\author{M. Barranco}
\affiliation{Departament E.C.M., Facultat de F\'{\i}sica, Universitat
de Barcelona, E-08028, Spain}

\author{M. Pi}
\affiliation{Departament E.C.M., Facultat de F\'{\i}sica, Universitat
de Barcelona, E-08028, Spain}

\begin{abstract}

We have experimentally studied the electronic  $3p\leftarrow 3s$
excitation of Na atoms attached to $^3$He droplets by means of
laser-induced fluorescence as well as beam depletion spectroscopy. From
the similarities of the spectra (width/shift of absorption lines) with
these of Na on $^4$He droplets, we conclude that sodium atoms reside in
a ``dimple'' on the droplet surface. The experimental results
are supported by Density Functional calculations at zero temperature,
which confirm the surface location of sodium on $^3$He droplets, and
provide a microscopic description of the ``dimple'' structure.


\end{abstract}

\date{\today}
\maketitle

Detection of laser-induced fluorescence (LIF) and beam depletion (BD)
signals upon laser excitation provides a sensitive spectroscopic
technique to investigate electronic transitions of
chromophores attached to $^4$He nanodroplets \cite{stienke2}. While
most of atomic and molecular dopants migrate to the center of the
droplet, alkali atoms (and alkaline earth atoms to some extent
\cite{Sti:1997b}) have been found to reside on the surface of $^4$He
droplets, as evidenced by the much narrower and less shifted spectra
when compared to those found in bulk liquid $^4$He
\cite{scoles,ernst1,stienke3,Ernst:2001a}. This result has been
confirmed by  Density Functional (DF) \cite{anci1} and Path Integral
Monte Carlo (PIMC) \cite{nakayama} calculations, which predict surface
binding energies of a few Kelvin, in agreement with the measurements of
detachment energy thresholds using the free atomic
emissions \cite{KKL}. The surface of liquid $^4$He is only slightly
perturbed by the presence of the impurity, which produces a ``dimple''
on the underlying liquid. The study of these states can thus provide
useful information on surface properties of He nanodroplets
complementary to that supplied by molecular-beam scattering experiments
\cite{Dal98,har01}.

Although the largest amount of work has been devoted to the study of
pure and doped $^4$He nanodroplets -see \cite{toennies,kwon} and
Refs.~therein-, the only neutral Fermi systems  capable of being
observed as bulk liquid and droplets are made of $^3$He atoms, and for
this reason they have  also attracted the interest of experimentalists
and theoreticians \cite{harms,har01,panda,TF,gar98,gua00}. We recall
that while $^4$He droplets, which are detected at an experimental
temperature ($T$) of $\sim$ 0.38\, K, are superfluid, these containing
only $^3$He atoms, even though detected at a lower $T$ of $\sim$
0.15\,K, do not exhibit superfluidity \cite{grebenev}.

The behavior of molecules in He clusters is especially appealing. In
particular, probes at the surface of the droplets are desirable because
they allow to investigate the liquid--vacuum interface as well as
droplet surface excitations. The latter are of interest in the
comparison of the superfluid vs. normal fluid behavior, particularly
because the Bose--Einstein condensate fraction has been calculated to
approach 100\% on the surface of $^4$He \cite{Griffin:1996}. Small
$^3$He drops are difficult to detect since, as a consequence of the
large zero-point motion, a minimum number of atoms is needed to produce
a selfbound drop \cite{panda}. Microscopic calculations of $^3$He
droplets are scarce, and only concern the ground state structure
\cite{panda,gua00}. Ground state properties and collective excitations
of $^3$He droplets doped with some inert atoms and molecular impurities
have been addressed within the Finite Range Density Functional (FRDF)
theory \cite{gar98}, that has proven to be a valuable alternative to
Monte Carlo methods, which are notoriously difficult to apply to Fermi
systems.
Indeed, a quite accurate
description of the properties of inhomogeneous liquid $^4$He at $T=0$ has
been obtained within  DF theory \cite{prica}, and a similar approach has
followed for $^3$He (see \cite{gar98,her02} and Refs.~therein).

The experiments we report have been performed in a helium droplet
machine used earlier for LIF and BD studies, and is described
elsewhere \cite{Sti:1997b}. Briefly, helium gas is expanded under
supersonic conditions from a cold nozzle forming a beam of droplets
traveling freely under high vacuum conditions. The droplets are doped
downstream employing the pick-up technique: in a heated scattering
cell, bulk sodium is evaporated in such a way that, on average, a
single metal atom is carried by each droplet. LIF absorption spectra of
doped droplets are recorded upon electronic excitation using a
continuous wave ring-dye laser and detection in a photo multiplier
tube (PMT).  Since electronic excitation of alkali-doped helium droplets is
eventually followed by desorption of the chromophore, BD spectra can be
registered by a Langmuir-Taylor surface ionization detector
\cite{Sti:2000b} housed in a separate
ultra-high vacuum chamber. Phase-sensitive
detection with respect to the chopped laser or droplet beam was used.
For that reason the BD signal (cf.~Fig.~\ref{exp_spectra}), i.e.~a
decrease in intensity, is directly recorded as a positive yield. For
these experiments, a new droplet source was built to provide the
necessary lower nozzle temperatures to condense $^3$He droplets.
Expanding  $P_0=$ 20\,bar of helium gas through a nozzle 5\,$\mu$m in
diameter, we now can establish temperatures down to 7.5\,K using a
two-stage closed cycle refrigerator (Sumitomo Heavy Industries, Model:
RDK-408D). In this way, without needing any liquid helium or nitrogen
for pre-cooling or cold shields, stable beam conditions can be utilized
over several days.

\begin{figure}
\resizebox{0.75\columnwidth}{!}{\includegraphics{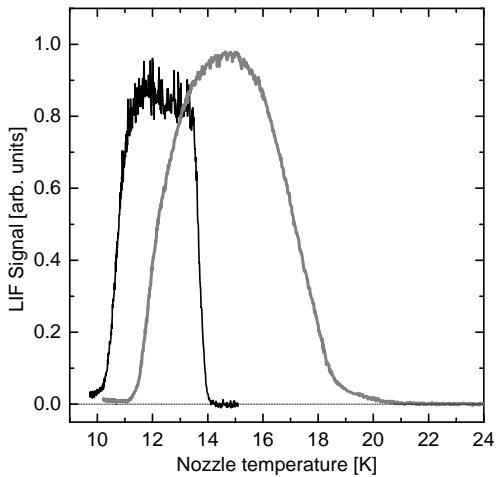}}
\caption{Laser-induced fluorescence signal as a function of nozzle
temperature forming $^3$He droplets (black) in comparison to $^4$He
droplets (grey). In both runs a stagnation pressure of 20\,bar was
used; nozzle diameter was 5\,$\mu$m. Normalization is such that the
plot gives the correct relative intensities.} \label{sourceconditions}
\end{figure}

\begin{figure}
\resizebox{0.75\columnwidth}{!}{\includegraphics{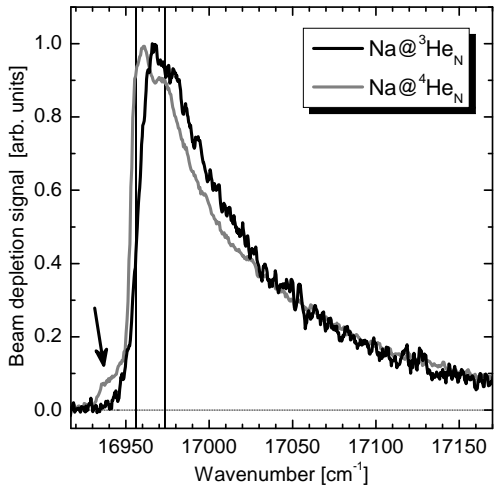}}
\caption{Beam depletion spectra of Na atoms attached to $^3$He/$^4$He
nanodroplets. The vertical lines indicate the positions of the two
components of the Na gas-phase  $3p\leftarrow 3s$
transition.}\label{exp_spectra}
\end{figure}

Fig.~\ref{sourceconditions} compares the number of Na-doped $^3$He
vs.~$^4$He droplets expanding 20\,bar helium as a function of the
nozzle temperature $T_0$. The absolute number densities in the maxima
of both distributions are quite similar. For the spectroscopic
measurements presented in the following, we set $T_0 = 11$\,K for
$^3$He, and $T_0 = 15$\,K for $^4$He. These conditions are expected to
result in comparable mean cluster sizes around 5000 atoms per droplet
\cite{Toe:unpublished,har01}. In Fig.~\ref{exp_spectra} the absorption
spectrum of Na atoms attached to $^3$He nanodroplets is shown in
comparison to Na-doped $^4$He droplets. We present here the BD spectra
because they do not contain the strong fluorescence background lines of
free Na atoms which cover the crucial steep increase of the droplet
spectrum. Moreover, LIF does not always represent the total absorption
spectrum because it relies on the emission of a photon in the spectral
range of the PMT. Hence, absorption processes followed by either
radiationless decay or emission of photons in the infrared spectral
region are suppressed. The latter has been observed in LIF spectra
where alkali--helium exciplexes form upon excitation of alkali atoms on
the surface of $^4$He droplets \cite{stienke3,Lehmann:2000c}. In our
experiment Na$^3$He exciplexes are formed in the same way as their
Na$^4$He counterparts. This became immediately obvious because we were
able to discriminate the corresponding red-shifted emission
intensities. Furthermore, we followed directly the Na$^3$He formation
in real-time applying femtosecond techniques. These results will be
presented in a forthcoming paper \cite{Sti:unpublisheda}. However, as
far as the measured absorption spectra of Na$@^3$He$_N$ are concerned,
the LIF data are very well in accord with the BD absorption.

The outcome of the spectrum of Na attached to $^3$He nanodroplets is
very similar to the spectrum on $^4$He droplets, which is remarkable
for two apparently very different fluids -one normal and the other
superfluid-, and confirms immediately the surface location: embedded
atoms are known to evolve large blue-shifts of the order of a couple of
hundreds of wavenumbers and much more broadened absorption lines, like
e.g.~observed experimentally in bulk helium \cite{Takahashi:1993}. The
blue shift is a consequence of the repulsion of the helium environment
against the spatially enlarged electronic distribution of the excited
state (``bubble effect''). The interaction towards the $^3$He droplets
appears to be slightly enhanced, evidenced by the small extra blue
shift of the spectrum compared to the $^4$He spectrum. In a simple
picture this means that more helium atoms are contributing or, in other
words, a more prominent dimple interacts with the chromophore. The
upper halves of the spectra come out almost identical, when shifting
the $^3$He spectrum 8\,cm$^{-1}$ to lower frequencies. The increase in
width (FWHM) of the spectrum is only 7\% for $^3$He droplets. Taking
into account the narrowing of the absorption line for small helium
droplets \cite{scoles} and the uncertainty in the $^3$He droplet size
distribution, this difference might not even be significant. Although
the width of the absorption might be hard to interpret, the blue shift
is a clear indication that the ``dimple'' is more pronounced than in
$^4$He droplets. Regarding the substructure of the line, the only
notable exception is the absence of the red-shifted shoulder, which is
observed in the case of $^4$He and marked with an arrow in
Fig.~\ref{exp_spectra}. This feature, which is even more pronounced in
the absorption of Li-doped $^4$He droplets \cite{scoles} has not been
interpreted yet.  The shift with respect to the maximum of the
absorption line is $\approx\,20\,$cm$^{-1}$, too high in energy to be
attributed to an excited compressional or surface mode of the droplet
at 0.38 K \cite{gar98,Chi95}.

\begin{figure} 
\resizebox{\columnwidth}{!}{\includegraphics{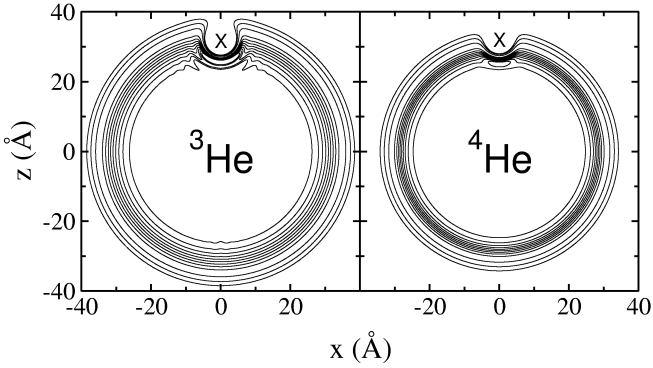}}
\caption{Equidensity lines in the $x-z$ plane showing the stable state
of a Na atom (cross) on a He$_{2000}$ droplet. The 9 inner lines
correspond to densities $0.9 \rho_0$ to $0.1 \rho_0$, and the 3 outer
lines to $10^{-2} \rho_0$, $10^{-3} \rho_0$, and  $10^{-4} \rho_0$
($\rho_0=0.0163$ \AA$^{-3}$ for $^3$He, and 0.0218 \AA$^{-3}$ for
$^4$He). } \label{fig3}
\end{figure}

FRDF calculations at $T=0$ confirm the picture emerging from the
measurements just reported, i.e.~the surface location of Na on $^3$He
nanodroplets causing a more pronounced ``dimple''  than in $^4$He
droplets. We have investigated the stable configurations of a sodium
atom on both $^3$He and $^4$He clusters of different sizes. The FRDF's
used for $^3$He and $^4$He are described in \cite{bar97}, with the
changes introduced in \cite{may01}. The large number of $^3$He atoms we
are considering allows to use the extended Thomas-Fermi approximation
\cite{TF}. The minimization of the energy DF's with respect to density
variations, subject to the constraint of a given number of He atoms
$N$, leads to Euler-Lagrange equations whose solution give the
equilibrium particle densities $\rho ({\bf r})$. These equations have
been solved as indicated in \cite{Bar03}. The presence of the foreign
impurity is modeled by a suitable potential obtained by folding the
helium density with a Na-He pair potential $V_{He-Na}$. Since the
adsorption properties of weakly interacting atoms on liquid He are
known to be very sensitive to the details of that potential, accurate
descriptions of the impurity-He interactions are mandatory for
quantitative predictions. We have used the potential proposed by Patil
\cite{patil}. Potential energy curves describing the He-alkali
interaction have been calculated by  ab-initio methods \cite{nakayama},
and found to agree very well with the Patil potential.

Fig.~\ref{fig3} shows the equilibrium configuration for a Na atom
adsorbed onto He$_{2000}$ clusters. For a given $N$, the size of the
$^3$He$_N$ droplet is larger than that of the $^4$He$_N$ droplet -an
obvious consequence of the smaller $^3$He saturation density-.
Comparison with the stable state on the $^4$He$_{2000}$ cluster  shows
that, in agreement with the experimental findings presented before, the
``dimple" structure is  more pronounced in the case of $^3$He, and that
the Na impurity lies {\it inside} the surface region for $^3$He and
{\it outside} the surface region for $^4$He \cite{note}. We attribute
this to the lower surface tension of $^3$He (0.113 K/{\rm \AA}$^2$) as
compared to that of $^4$He (0.274 K/{\rm \AA}$^2$), which also makes
the surface thickness of bulk liquid and droplets larger for $^3$He
than for $^4$He \cite{Dal98,har01}. The Na-droplet equilibrium
distance, here defined as the `radial' distance  between the impurity
and the point where the density of the pure drop would be $\rho \sim
\rho_0/2$ ($\rho_0$ being the saturation density of bulk liquid He), is
$R \sim 1.1$ \AA\, for $^3$He, and $R \sim 3.6$ \AA\, for $^4$He. For
the larger droplets we have studied, $R$ is nearly $N$-independent. A
related quantity is the deformation of the surface upon Na adsorption,
which can be characterized \cite{anci1} by the dimple depth, $\xi$,
defined as the difference between the position of the dividing surface
at $\rho \sim \rho_0/2$, with and without impurity, respectively. We
find $\xi = 4.4$ \AA \,for $^3$He, and $\xi = 1.8$ \AA $\,$ for $^4$He.

\begin{figure} 
\resizebox{0.9\columnwidth}{!}{\includegraphics{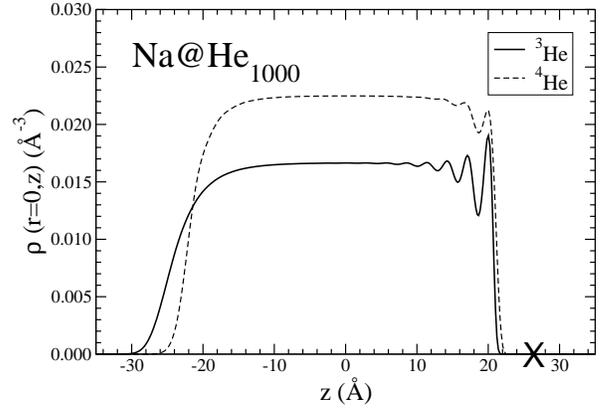}}
\caption{Density profiles along a line connecting the impurity to the
center of the cluster showing the equilibrium configuration of a Na
atom (cross) on He$_{1000}$ nanodroplets.} \label{fig4}
\end{figure}

Fig.~\ref{fig4} shows the  density profiles for Na$@$He$_{1000}$. Note
the more diffuse liquid-vacuum interface for the $^3$He droplet far
from the impurity, and the occurrence of more marked density
oscillations (with respect to the $^4$He case) where the softer $^3$He
surface is compressed by the adsorbed Na atom. 
Our calculations thus
provide a detailed picture of the He structure around the
impurity, which is an essential ingredient for any line-shape calculation of
the main electronic transitions in the adatom \cite{scoles}.

\begin{figure} 
\resizebox{0.9\columnwidth}{!}{\includegraphics{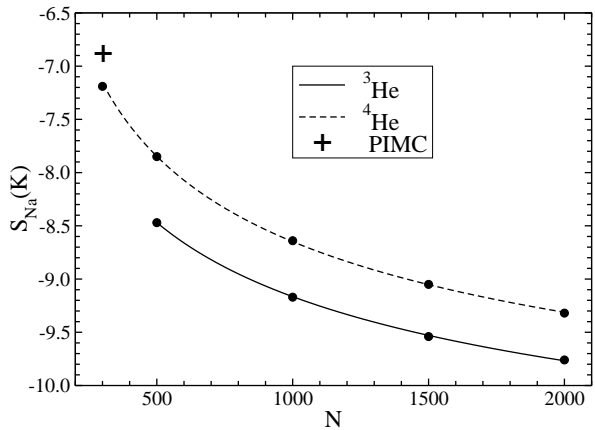}} \caption{ Na
solvation energy as a function of the cluster size.
The cross is the PIMC result of \cite{nakayama}.
} \label{fig5}
\end{figure}

Solvation energies defined as 
\begin{equation}
 S_{Na}=E({\rm Na}@{\rm He}_N)-E({\rm He}_N) 
\end{equation}
are shown in Fig.~\ref{fig5}. To compare with ``exact'' PIMC
result \cite{nakayama}, we have calculated Na$@^4$He$_{300}$. Part of
the small difference between the two values has to be attributed to our
neglecting of the zero-point energy of Na. 

We have fitted the
$S_{Na}$'s to a mass formula of the kind
\begin{equation}
S_{Na}(N)= S_0+ \frac{S_1}{N^{1/3}} + \frac{S_2}{N^{2/3}}
\end{equation}
and have found $S_0=-12.1 (-12.5)$, $S_1=-28.3 (-31.6)$, and $S_2=37.3
(37.6)$ K for $^4$He($^3$He). The lines in Fig.~\ref{fig5} are the
result of the fit. The value $S_{Na} \sim -12$ K has been obtained
within FRDF theory for Na adsorbed on the {\it planar} surface of
$^4$He \cite{anci1}, which would correspond to the $N=\infty$ limit in
Fig. \ref{fig5}.

In conclusion, the presented experimental and theoretical results show
that alkali adsorption on $^3$He droplets occurs in very much the same
way as in the case of $^4$He, i.e. the adatom is located on the surface
in a slightly more pronounced ``dimple". Our calculations provide a
microscopic description of the liquid structure around the impurity,
which is a necessary input for any line-shape calculation as well as
for the understanding of dynamical processes which already have been
observed in time-dependent experiments \cite{Sti:unpublisheda}.

We thank Flavio Toigo for useful comments. This work has been performed
under grants MIUR-COFIN 2001 (Italy), BFM2002-01868 from DGI (Spain),
and 2001SGR-00064 from Generalitat of Catalunya. Financial support by
the Deutsche Forschungsgemeinschaft (Germany) is gratefully
acknowledged.


\begin{thebibliography}{99}
\bibitem{note}
The surface region is usually defined as that comprised between
the radii at which $\rho=0.1 \rho_0$ and
$\rho=0.9 \rho_0$, where $\rho_0$ is the He
saturation density \cite{Dal98,har01,TF}.
\end{thebibliography}
\end{document}